\documentclass{elsarticle}
\usepackage{amsmath}
\usepackage{graphicx}
\input{tcilatex}

\begin{document}

\begin{frontmatter}
\title{    Limits and Confidence Intervals in the Presence of Nuisance Parameters}
\author[wolf]{Wolfgang A. Rolke}
\address[wolf]{Department of Mathematics, University of Puerto Rico - Mayag\"{u}ez, Mayag\"{u}ez, PR 00681, USA, 
\newline Postal Address: PO Box 5959, Mayag\"{u}ez, PR 00681, 
\newline Tel: (787) 255-1793, Email: wolfgang@puerto-rico.net}
\author[ang]{Angel M. L\'{o}pez}
\address[ang]{Department of Physics, University of Puerto Rico - Mayag\"{u}ez, Mayag\"{u}ez, PR 00681, USA}
\author[jan]{Jan Conrad}
\address[jan]{PH-Department, CERN, CH-1211, Geneva 23, Switzerland.}

\begin{abstract}
   We study the frequentist properties of confidence intervals computed by 
the method known to statisticians as the Profile Likelihood.  It is seen that the coverage of 
these intervals is surprisingly good over a wide range of possible 
parameter values for important classes of problems, in particular whenever there 
are additional nuisance parameters with  statistical or systematic errors. 
Programs are available for calculating these intervals.
\end{abstract}
\end{frontmatter}\newpage

\section{Introduction}

The calculation of confidence intervals (or setting of limits) on a
parameter of a theory is one of the most important problems an experimental
physicist can face. In the frequentist approach which we follow here, the
main property which confidence intervals have to fulfill is to have
coverage. A method is said to yield a $100(1-\alpha )\%$ confidence interval
if, were the experiment to be repeated many times, the resulting intervals
would include (or cover) the true parameter at least $100(1-\alpha )\%$ of
the time, no matter what the true parameter is. Using a construction method
due to Neyman \cite{Neyman:1937a}, Feldman and Cousins \cite{Feldman and
Cousins} in 1998 found confidence intervals for the case of one nuisance
parameter when its value is known exactly. An alternative method used widely
in high energy physics prior to the publication of Feldman and Cousins is to
extract confidence intervals by finding the points where the -2log
likelihood function increases by a factor defined by the required confidence
level ($\ln {\mathcal{L}}+\frac{1}{2}$ method). This method is also well
known in Statistics as the large-sample approximation to the likelihood
ratio test statistic. One major drawback of this method is that, because it
is based on a large-sample theory, its performance for small samples is in
question. In fact, it is known that the $\ln {\mathcal{L}}+\frac{1}{2}$
method has under-coverage in certain circumstances. However, it does not
suffer from the few practical drawbacks of Feldman and Cousins. Firstly, it
can be generalized to the case where there are many parameters of interest.
Secondly (and more importantly here) it is easily adapted to treat problems
with several nuisance parameters which are not known exactly. In this note
we will combine the $\ln {\mathcal{L}}+\frac{1}{2}$ method with what is
known in Statistics as the profile likelihood method in which the
multi-dimensional likelihood function is reduced to a function that only
depends on the parameter of prime interest. The combination of the profile
likelihood approach with the $\ln {\mathcal{L}}+\frac{1}{2}$ extraction of
confidence intervals is well known to particle physicists from the program
MINUIT \cite{James and Roos}, \cite{James} and to astrophysicists from
SERROR \cite{Lampton} and is discussed at
http://www.sr.bham.ac.uk/asterix-docs/Programmer/Source/Algorithms/serror.html.

For the problem of setting confidence limits for the signal rate in the
presence of background which is estimated from data sidebands or Monte
Carlo, this approach has previously been shown in Rolke and L\'{o}pez \cite%
{Rolke and Lopez} to have good coverage. In this note the method will be
generalized to the problem of a signal with a Poisson distribution, a
background with either a Poisson or a Gaussian distribution and an
efficiency with either a Binomial or a Gaussian distribution. We will
establish the domain of validity, enabling comparison with other methods. As
we will show, this method, together with some minor adjustments, has very
good coverage even in cases when the parameters lie close or at the physical
boundaries.

Although this paper, and the corresponding routines, only deal with the
specific problems outlined above, the results show that the method of
profile likelihood is a viable technique for dealing with nuisance
parameters, and it should be useful for other problems as well.

\section{The Method}

\subsection{Profile Likelihood}

The basic idea of the profile likelihood is rather straightforward: assume
we have a probability model for our data which depends on parameters $%
\mathbf{\pi }=\left( \pi _{1},..,\pi _{k}\right) $ of interest to the
researcher but also on additional nuisance parameters $\mathbf{\theta }%
=\left( \theta _{1},..,\theta _{l}\right) $. If we denote the probability
mass function (or density) by $f(x|\mathbf{\pi },\mathbf{\theta })$ and we
have independent observations $\mathbf{X}=\left( X_{1},..,X_{n}\right) $,
then the full likelihood function is given by 
\begin{equation*}
L(\mathbf{\pi },\mathbf{\theta }|X)=\dprod\limits_{i=1}^{n}f(X_{i}|\mathbf{%
\pi },\mathbf{\theta })
\end{equation*}
If, as is often done in Statistics, we decide to base our inference on the
likelihood, our challenge is to eliminate the nuisance parameters.

A standard technique for constructing confidence intervals is to find a
corresponding hypothesis test, and then to invert that test. Here the
hypothesis test is $H_{0}:\mathbf{\pi =\pi }_{0}$ vs $H_{a}:\mathbf{\pi \neq
\pi }_{0}$ and a test can be based on the likelihood ratio test statistic
given by 
\begin{equation*}
\lambda (\mathbf{\pi }_{0}|X)=\frac{\sup \left\{ L(\mathbf{\pi }_{0},\mathbf{%
\theta }|X);\mathbf{\theta }\right\} }{\sup \left\{ L(\mathbf{\pi },\mathbf{%
\theta }|X);\mathbf{\pi },\mathbf{\theta }\right\} }
\end{equation*}
The supremum in the denominator is found over the full parameter space,
whereas the supremum in the numerator is found only over the subspace with $%
\mathbf{\pi =\pi }_{0}$. Notice that $\lambda $ is a function of $\mathbf{%
\pi }_{0}$ (and the data) only, but does not depend on the nuisance
parameters $\mathbf{\theta }$. In the context of nuisance parameters the
function $\lambda $ is also called the profile likelihood. One of the
standard results from Statistics (see for example Casella and Berger \cite%
{Casella-Berger}) is that $-2\log \lambda $ converges in distribution to a
chi-square random variable with $k$ degrees of freedom. This in fact is the
theoretical basis for the $\ln {\mathcal{L}}+\frac{1}{2}$ method for
extracting limits from the likelihood function.

We illustrate the method of profile likelihood using the example of a search
for a rare decay where the expected background is known only approximately.
We will need the following notation. Assume that we observe $x$ events in a
suitably chosen signal region and a total of $y$ events in the background
region. Here the background region can be chosen fairly freely and need not
be contiguous. Furthermore, the probability that a background event falls
into the background region divided by the probability that it falls into the
signal region is denoted by $\tau $. For example, if we use two background
regions of the same size as the signal region and assume the background
distribution is flat we have $\tau =2$. If the background rate is estimated
from Monte Carlo, $\tau $\ is the size of the Monte Carlo sample relative to
the size of the data sample. Then a probability model for the data is given
by 
\begin{equation*}
X\sim Pois(\mu +b),\qquad Y\sim Pois(\tau b)
\end{equation*}
where $\mu $ is the signal rate, $b$ is the background rate and $Pois$ is
the usual Poisson distribution. We will use large caps letters $X,Y$ to
denote random variables and small caps letters $x,y$ to denote realizations
(observed values) of these random variables. We can assume $X$ and $Y$ to be
independent and so 
\begin{equation*}
f(x,y|\mu ,b)=\frac{(\mu +b)^{x}}{x!}e^{-(\mu +b)}\cdot \frac{(\tau b)^{y}}{%
y!}e^{-\tau b}
\end{equation*}

The likelihood function is given by $L(\mu ,b|x,y)=f(x,y|\mu ,b)$.\
Maximizing over both $\mu $ and $b$ we find the usual maximum likelihood
estimators $(\widehat{\mu },\widehat{b})=(x-y/\tau ,y/\tau )$. Fixing $\mu $
and maximizing over b alone yields 
\begin{equation*}
\hat{b}(\mu )=\frac{x+y-(1+\tau )\mu +\sqrt{\left( x+y-(1+\tau )\mu \right)
^{2}+4(1+\tau )y\mu }}{2(1+\tau )}
\end{equation*}%
For other models it may not be possible to find $\hat{b}(\mu )$
analytically, in which case numerical methods need to be used. Now the
profile likelihood function is given by 
\begin{equation*}
\lambda (\mu |x,y)=\frac{L(\mu ,\hat{b}(\mu )|x,y)}{L(\widehat{\mu },%
\widehat{b}|x,y)}
\end{equation*}%
and according to the theory $-2\log \lambda $ has an approximate $\chi ^{2}$
distribution with $1$ degree of freedom.

For more details on the likelihood ratio test statistic see Casella and
Berger \cite{Casella-Berger}. For information on the profile likelihood see
Bartlett \cite{Barlett}, Lawley \cite{Lawley} and Murphy and Van Der Vaart 
\cite{Murphy and Van Der Vaart}.

For the studies conducted here, the maximizations leading to $-2\log \lambda 
$ have mostly been done analytically, but we have also verified that the
program MINUIT (which does this numerically) produces the same results when
using MINOS errors if $x\geq y/\tau $. The case of $x<y/\tau $ is discussed
in detail in section 2.3.

There are a number of refinements known in Statistics for the profile
likelihood. We had to make a judgement call here: keep the method very
simple and easy to implement for different types of problems, or include
further refinements which might improve the method. Given that it works as
well as it does any improvement would be very slight, at least for the
problems studied in detail. We consider it a definite strength of our method
that it should be very easy to apply to other problems, and this advantage
would be lost if we advocated such refinements.

\subsection{The Treatment of Efficiency and Systematic Errors}

The general nature of the profile likelihood technique for dealing with
nuisance parameters can be illustrated by considering several modifications
and extensions of the problem as laid out in the previous paragraphs. For
example, say we want to include the efficiency $e$ into our limits. Assume
that we are Monte Carlo limited and therefore have to deal with the error in
the efficiency estimate. Specifically, say we run $m$ events through our
Monte Carlo (without background) and find $z$ events surviving.\ Then we can
model the efficiency $Z$ as a binomial random variable and find the complete
model to be 
\begin{equation*}
X\sim Pois(e\mu +b),\qquad Y\sim Pois(\tau b),\qquad Z\sim Bin(m,e)
\end{equation*}
where $Bin$ is the binomial distribution. To find the profile likelihood we
fix $\mu $ and differentiate the loglikelihood: 
\begin{equation*}
\frac{\partial }{\partial b}\log l(\mu ,b,e|x,y,z)=\frac{x}{e\mu +b}-1+\frac{%
y}{b}-\tau \doteq 0
\end{equation*}

\begin{equation*}
\frac{\partial }{\partial e}\log l(\mu ,b,e|x,y,z)=\frac{x}{e\mu +b}-\mu +%
\frac{z}{e}-\frac{m-z}{1-e}\doteq 0
\end{equation*}
This system of nonlinear equations can not be solved analytically but we can
do so numerically, and again we have the profile likelihood curve as a
function of the signal rate $\mu $ alone.

As a second example, suppose that the background and the efficiency are
better modeled as Gaussians rather than using the Poisson and the Binomial,
for example, to allow the inclusion of systematic errors. Then we have the
probability model 
\begin{equation*}
X\sim Pois(e\mu +b),\qquad Y\sim N(b,\sigma _{b}),\qquad Z\sim N(e,\sigma
_{e})
\end{equation*}%
where $N$ indicates the Gaussian (or normal) distribution and $\sigma _{b}$
and $\sigma _{e}$ are the standard deviations or errors on the estimates of $%
b$ and $e$, respectively. Now we find the derivatives of the loglikelihood
to be 
\begin{equation*}
\frac{\partial }{\partial b}\log l(\mu ,b,e|x,y,z)=\frac{x}{e\mu +b}-1+\frac{%
(y-b)}{\sigma _{b}}\doteq 0
\end{equation*}

\begin{equation*}
\frac{\partial }{\partial e}\log l(\mu ,b,e|x,y,z)=\frac{x}{e\mu +b}-\mu +%
\frac{(z-e)}{\sigma _{e}}\doteq 0
\end{equation*}
This system can actually be solved analytically.

All combinations of the above models, such as the background modeled as a
Poisson and the efficiency modeled as a Gaussian as well as the cases where
one or the other or both are known without error, are equally easily treated.

\subsection{Extracting Limits}

As was discussed above $-2\log \lambda $ has approximately a chi-square
distribution, and this can be used to extract limits. Other methods also
exist, for example, using information from the second derivative of the log
profile likelihood. However, for the specific problem of rare decays, this
does not yield a method with correct coverage.

Figure 1 shows $-2\log \lambda $ for the case $x=8$, $y=15$, $\tau =5.0$. To
find a $100(1-\alpha )\%$ confidence interval we start at the minimum, which
of course is at the usual maximum likelihood estimator, and then move to the
left and to the right to find the points where the function increases by the 
$\alpha $ percentile of a $\chi ^{2}$ distribution with $1$ degree of
freedom. For example, if we want to find a $90\%$ confidence interval, the
increase will be $2.706.$

Some care must be taken in the cases where the maximum likelihood estimator
is small or negative. If it is positive but sufficiently small, $-2\log
\lambda $ might not increase enough to the left within the physical region ($%
\mu \geq 0$). In that case, we set the lower limit to zero. (All our
techniques produce physically valid limits.)

In the cases where fewer events are observed in the signal region than are
expected from background, the log profile likelihood curve, just like the
regular log likelihood, is no longer parabolic. Rolke and L\'{o}pez \cite%
{Rolke and Lopez} dealt with this problem by using a hypothesis test based
on the null hypothesis $H_{0}:\mu =\mu _{0},b=b_{0}$, deriving the
corresponding two-dimensional acceptance region and then finding the values
of $\mu $ where the profile likelihood $\left( \mu ,\hat{b}(\mu )\right) $
enters and leaves the acceptance region. Rolke and L\'{o}pez \cite{Rolke and
Lopez} also extended this method to a problem with two nuisance parameters
but this becomes computationally very demanding because it requires
searching for boundary points of a highly irregular object in two (or more)
dimensions.

In this paper we consider two methods for treating these cases. In the
first, referred to as the unbounded likelihood method, we proceed just as
described above. This is possible because for the problems studied here we
can always find the maximum likelihood estimators analytically and therefore
the value of $-2\log \lambda $ at that point. The maximum likelihood
estimator for the signal rate is negative but we set the lower limit to
zero. As an example consider the left panel of figure 2. Here we have $x=2$, 
$y=15$, $\tau =5.0$, so the maximum likelihood estimator of the signal rate
is $\widehat{\mu }=-1.0$. Using the method of unbounded likelihood we find a 
$95\%$ upper limit of $3.35$.

In the most extreme case when we observe many fewer events than are expected
from background alone, the profile likelihood curve at $\mu =0$ might
already be higher than the increase from the minimum. Rather than quoting a
negative (and non-physical) upper limit we will find the upper limit by
increasing the value of $x$ by $1$ until we find the first positive upper
limit. This will yield a legitimate upper limit because our limits are
monotonically increasing in $x$.

A special treatment is also necessary for the cases $x=0$ and/or $y=0$. This
is due to the fact that here the loglikelihood function is linear and
therefore does not have a minimum, that is, the maximum likelihood estimator
does not exist. We solve this problem by computing the "neighboring" limits (%
$x=1,2$ and/or $y=1,2$) and doing a linear extrapolation. This is a
reasonable \textquotedblleft adjustment\textquotedblright\ because the
limits are nearly linear in $x$ and $y$, and as we shall see in the next
section the resulting confidence intervals have good coverage properties.

A second possible solution to the problem of fewer events in the signal
region than are expected from background is to use the physical limits on
the parameters. So instead of using the increase from the maximum likelihood
estimator $\widehat{\mu }$ we instead use the increase from the point $\mu
=0 $ if $\widehat{\mu }<0$. This is equivalent to finding the minimum using
MINUIT with a lower bound for the signal rate of $0$. We will refer to this
as the method of bounded likelihood. It is illustrated in the right panel of
figure 2. Using this method we find a $95\%$ upper limit of $3.6$.

The adjustment for the cases $x=0$ and/or $y=0$ described above is also
necessary for the bounded likelihood method because without it the method
does not have coverage; for example, a nominal $90\%$ confidence interval
has a\ true coverage of less than $68\%$ for some parameter values.

The adjustments described here are rather ad-hoc, and their only theoretical
justification is that they make the limits larger, therefore can not make
the coverage worse. Their practical justification is of course the fact that
the resulting limits have coverage. Such ad-hoc adjustments are not at all
unusual in Statistics. Certainly one would prefer a method that deals
uniformly with all possible values in data space, but such methods are
actually very rare.

The treatment here deals only with the statistical aspect of the problem of
having fewer events in the signal region than are expected from background.
In such a situation we recommend to also quote the experimental sensitivity
which is defined as the mean upper limit for an ensemble of experiments with
the observed background and efficiency levels but no signal. In the case
where there are fewer events in the signal region than expected from
background, the experimental sensitivity gives a better idea of the
experimental precision than the upper limit. An even better way to proceed
in such cases is to use a "bootstrap procedure", described in Rolke and
Lopez \cite{Rolke and Lopez 2}, to reduce the effects of statistical
fluctuations in the background. However, such considerations are outside the
scope of this paper. Sensitivity determinations and the bootstrap procedure
require what we provide here, a method for calculating limits given any set
of observations.

Finally, the method does not return limits for some cases where the Gaussian
model for the efficiency is suspect, that is, cases where $z/\sigma _{e}$ is
small. One example is shown in figure 3 where we have $x=5$, $y=2.5$, $%
\sigma _{b}=0.4$ and $z=0.2$. Then for $\sigma _{e}=0.1$ the curve rises
sufficiently so we can quote an upper limit ($85.9$) but if $\sigma _{e}=0.15
$ it does not. It is impossible to give a precise value for the ratio below
which no limits are returned since it also depends on the values of the
observations and the confidence level desired. In practice, if the method
returns limits, coverage studies (discussed in Section 3) show them to be
reliable. If it does not, alternative models for the efficiency distribution
will need to be considered.

\section{Performance of this Method}

In the case of confidence intervals, performance means first of all
\textquotedblleft coverage\textquotedblright , that is, a nominal $90\%$
confidence interval should cover the true value of the parameter at least $%
90\%$ of the time for all parameter values. Because in the specific problem
discussed here at least one of the observations has a discrete distribution,
it is not possible to achieve the nominal coverage for all parameter values.
Some overcoverage is unavoidable, although less overcoverage is of course
preferable and is a measure of the quality of the method.

Coverage studies for the case of a Poisson model for the signal and a
Poisson model for the background have previously been published in Rolke and
L\'{o}pez \cite{Rolke and Lopez}. For the case of an added efficiency
modeled as a Binomial (discussed above), consider the following coverage
study: We have $\tau =3.5$, the efficiency is $e=0.85$ and $m=100$. We vary
the signal rate $\mu $ from $0$ to $10$ in steps of $0.1$ and the background
rate $b$ from $0$ to $10$ in steps of $2$. For each of these $600$
combinations of the parameters we find the true coverage of nominal $90\%$
confidence intervals based on $10000$ Monte Carlo runs and for each of the
methods described above. The results are shown in figure 4. As a second
example we model both the background and the efficiency as Gaussians, with $%
\sigma _{b}=0.5$ and $\sigma _{e}=0.075$. Again we have $e=0.85$ and vary $%
\mu $ and $b$ as above. The results are shown in figure 5.

In both figure 4 and figure 5 we included the case $b=0$ to show that our
method is capable of dealing even with such an extreme case with values at
the boundary of parameter space.

The coverage of the method is quite good, with only some very small
acceptable undercoverage due to the fact that we are using a large-sample
approximation. We have some overcoverage for small $\mu $ because the upper
limit can not be too small in this case but this is unavoidable. Finally we
see that these coverage graphs are considerably smoother with much less
overcoverage than those shown in Rolke and L\'{o}pez \cite{Rolke and Lopez}
for both the unified method by Feldman and Cousins as well as the method
described there. This is because of the higher randomness due to the extra
random variable $Z$ for the efficiency. The overcoverage of the unbounded
likelihood method is generally a little smaller than for the bounded
likelihood method. This is as expected because in the cases where they
differ the unbounded likelihood method yields lower upper limits than the
bounded likelihood method. Extensive coverage studies of over $25000$
parameter combinations and for all the models discussed above have shown
these results to be quite general.

In our coverage studies we do not include simulation runs for which our
method does not return limits. To have \textquotedblleft
coverage\textquotedblright\ is to make sure a fixed (and known) percentage
of published intervals actually contain the true parameter value. The cases
where no limit is returned will not lead to incorrect intervals but rather
to a reevaluation of the efficiency model. Such cases should occur very
rarely in practice and, in fact, this is so for the coverage studies in
figures 4 and 5. A coverage study where the percentage of such cases is
substantial is presented in table 1. Here we have $\mu =5$, $b=2.5$, $\sigma
_{b}=0.4$, $\sigma _{e}=0.1$ and we find $90\%$ confidence intervals. The
method has coverage for all values of the parameters, even the extreme value
where $e/\sigma _{e}=1$. Although it does overcover under these
circumstances, it does give conservative limits. Even for values of $%
e/\sigma _{e}$ as small as $3$ where the percentage of cases that do not
return a limit is $8\%$, the degree of overcoverage is small. However, the
Gaussian model should be reconsidered for such small ratios.

Table 1: Percentage of simulation runs that do not return limits and
coverage for cases where $e/\sigma _{e}$ is small:

\begin{tabular}{|l|l|l|}
\hline
$e$ $(e/\sigma _{e})$ & No Limits & Coverage \\ \hline\hline
$0.5$ $(5.0)$ & \multicolumn{1}{|r|}{$0.04\%$} & \multicolumn{1}{|r|}{$90\%$}
\\ \hline
$0.4$ $(4.0)$ & \multicolumn{1}{|r|}{$0.8\%$} & \multicolumn{1}{|r|}{$90\%$}
\\ \hline
$0.3$ $(3.0)$ & \multicolumn{1}{|r|}{$8\%$} & \multicolumn{1}{|r|}{$92\%$}
\\ \hline
$0.2$ $(2.0)$ & \multicolumn{1}{|r|}{$35\%$} & \multicolumn{1}{|r|}{$96\%$}
\\ \hline
$0.1$ $(1.0)$ & \multicolumn{1}{|r|}{$73\%$} & \multicolumn{1}{|r|}{$98\%$}
\\ \hline
\end{tabular}

In figure 6 we show the behavior of the limits as functions of the
uncertainties in background (left panel) and efficiency (right panel). The
limits are for the case $x=5$, $y=3$, $z=0.5$, and we model both the
background and the efficiency as Gaussians. In the left panel we vary the
uncertainty in the background rate from $0.0$ to $1.0$ with the uncertainty
in the efficiency fixed at $0.1.$ In the right panel we vary the uncertainty
in the efficiency from $0.0$ to $0.15$ with the uncertainty in the
background fixed at $0.75$. As we can see the behavior of the limits here is
what one expects: the larger the uncertainty the higher the limit. We also
see that the limits found using this method are self-consistent, that is, as
the uncertainty becomes small the limits smoothly approach a limiting value.

As the coverage graphs show, the unbounded likelihood method generally has
less overcoverage than the bounded likelihood method. This is due to the
fact that when they differ the upper limits of the unbounded likelihood
method are always smaller. This can be seen as an advantage and a reason to
prefer the unbounded likelihood method to the bounded likelihood method.

Based on our coverage studies we recommend to use the presented method for
Poisson statistics when the uncertainties in nuisance parameters can be
modeled as one of the seven cases we studied here. For problems other than
these the profile likelihood method should be useful as well, but then
additional coverage studies need to be carried out.

\section{Comparison with other methods}

An alternative to the method described in this paper is based on integrating
over the nuisance parameters, see for example Cousins and Highland \cite%
{Cousins Highland} and Conrad et. al. \cite{Conrad}. There the full
confidence interval construction is performed, but the probability density
function is obtained from folding the primary PDF with the one describing
the distribution of the nuisance parameters. Treating nuisance parameters in
this way is based on Bayesian statistics with a flat prior. The coverage of
this method has been studied in Tegenfeldt and Conrad \cite{Tegenfeld}. It
appears that treating nuisance parameters in a Bayesian way generally
introduces a modest amount of overcoverage, though it has to be emphasized
that this has been shown only for the assumed flat prior distribution. In
general so-called credible intervals found via Bayesian statistics do not
have the property of coverage, and the frequentist properties of such
intervals will have to be studied for each problem and each prior anew.

Feldman \cite{Feldman} suggested to use the profile likelihood in the ratio
ordering of the unified approach in an effort to extend that method to the
case of background uncertainty. Coverage studies for this method have not
been presented yet.

A standard method for finding errors, and therefore limits, is based on the
second derivative of the log likelihood function. For the problem described
here, though, this method can have a true coverage of less than $50\%$ for a
nominal $90\%$ confidence interval, and is therefore not a viable option.

As discussed above our method is very similar to MINUIT/MINOS; in fact for
the case $x>y/\tau $ the two yield identical limits. MINUIT/MINOS alone,
though, does not have coverage because it does not include the adjustments
described in section 2.3.

In figure 6 we have included the limits from Feldman and Cousins unified
approach in the case of zero uncertainty in the background rate. Those
limits are slightly different from ours. That different methods yield
different limits is very common in Statistics because the requirement of
coverage is in fact a rather general one and does not uniquely determine the
limits.

\section{Implementation}

A stand-alone FORTRAN routine for calculating these limits is available at
http://charma.uprm.edu/\symbol{126}rolke/publications.htm. It is also
available as TRolke which is part of ROOT \cite{Root}. Both routines also
calculate the experimental sensitivity described above. Finally, at least
for the cases $x>y/\tau $, one could use MINUIT/MINOS to carry out the
calculations, though in this case care needs to be taken to set the limits
on the parameters correctly.

It is to be hoped that the profile likelihood method yields good results
also in situations other than the ones discussed here. Because it is already
available as part of MINUIT, its implementation for different problems
should be quite straightforward. It needs to be emphasized, though, that the
profile likelihood method can not be assumed to yield good results in all
cases and that it might require some adjustments to the general method as we
have done here. It is therefore strongly recommended that a thorough check
of its performance be done whenever it is applied to a new problem. In the
case of setting limits, this means a coverage study as described above, at
least for the range of likely parameter values.

\section{Summary}

We have discussed the method of profile likelihood as a general treatment of
nuisance parameters within a frequentist framework. For the case of a
Poisson distributed signal with a background that has either a Poisson or a
Gaussian distribution and an efficiency that has either a Binomial or a
Gaussian distribution, we have carried out an extensive coverage study and
shown that the method yields confidence intervals with good coverage
throughout the parameter space, even at its boundaries.

\section{Acknowledgements}

J.C. would like to thank R. Brun for support with ROOT and G. Stefanini for
his patience. All authors would like to thank F. James for pointing out the
connection to MINUIT and for various discussions that have allowed us to
improve this paper considerably. This work was supported by the Division of
High Energy Physics (Grant DE-FG-97ER41045) of the US Department of Energy
and the University of Puerto Rico.

\section{Appendix}

\begin{figure}
	\centering
		\includegraphics[width=0.90\textwidth]{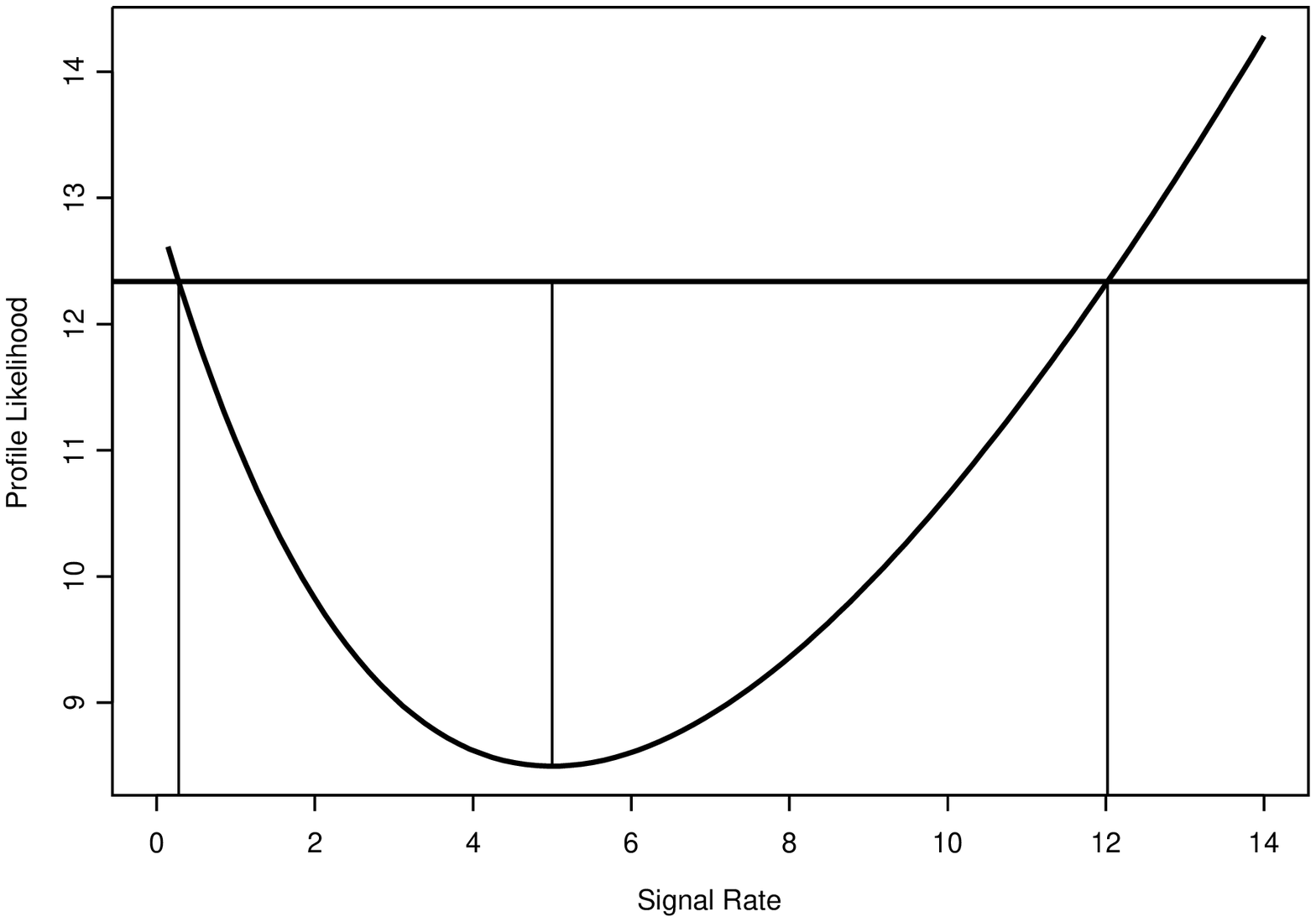}
	\caption{Example of the $-2\log 
\protect\lambda $ curve. This is the case $x=8$, $y=15$ and $\protect\tau %
=5.0$. We find the $95\%$ confidence interval to be$\ (0.28,12.02)$.}
	\label{fig:fig1}
\end{figure}

\begin{figure}
	\centering
		\includegraphics[width=0.90\textwidth]{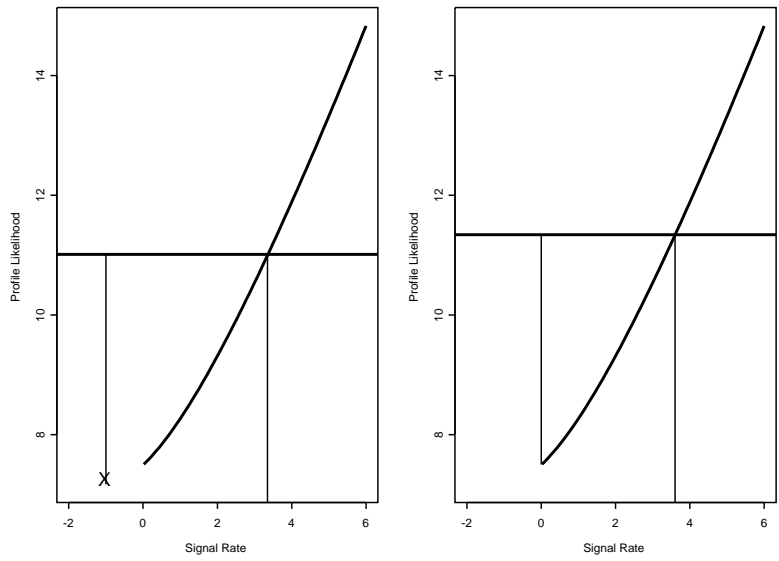}
	\caption{The case $x=2 $, $y=15$
and $\protect\tau =5.0$. In the left panel we use the unbounded likelihood
method and find a $95\%$ upper limit of $3.35.$ In the right panel using the
bounded likelihood method the $95\%$ upper limit is $3.6$.}
	\label{fig:fig2}
\end{figure}

\begin{figure}
	\centering
		\includegraphics[width=0.90\textwidth]{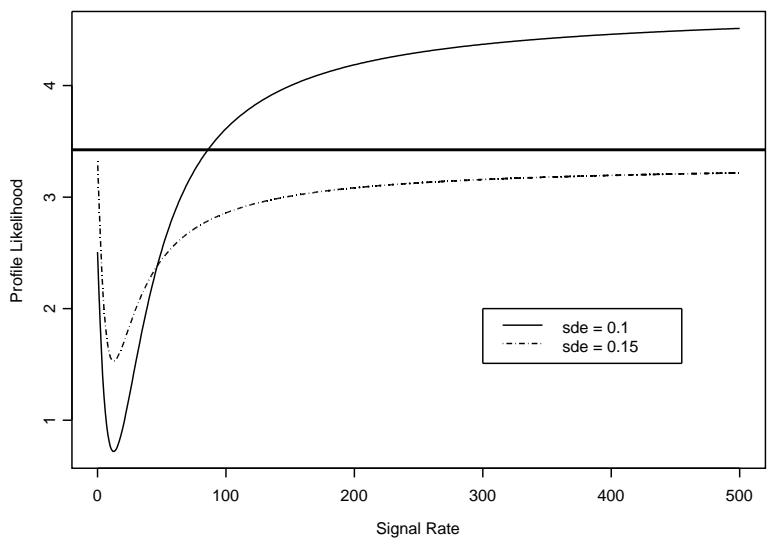}
	\caption{An illustration of why
the profile likelihood method does sometimes not return a limit. Here $x=5$, 
$y=2.5$, $\protect\sigma _{b}=0.4$ and $z=0.2$. If $\protect\sigma _{e}=0.1$
the curve moves above the required level ( for a 90\% confidence interval)
and the upper limit is $85.9$ but if $\protect\sigma _{e}=0.15$ the curve
stays below the level and no limit is found.}
	\label{fig:fig3}
\end{figure}

\begin{figure}
	\centering
		\includegraphics[width=0.90\textwidth]{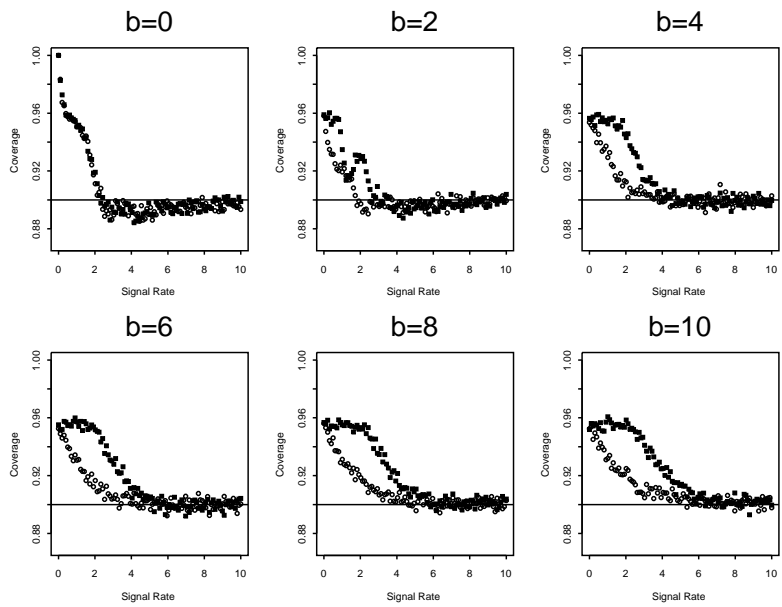}
	\caption{$90\%$ coverage graphs
when the signal and the background are modeled as Poisson and the efficiency
is modeled as a Binomial with $\protect\tau =3.5$, $e=0.85 $ and $m=100$.
The empty circles show the coverage using the unbounded likelihood method
and the solid squares show the coverage using the bounded likelihood method.}
	\label{fig:fig4}
\end{figure}

\begin{figure}
	\centering
		\includegraphics[width=0.90\textwidth]{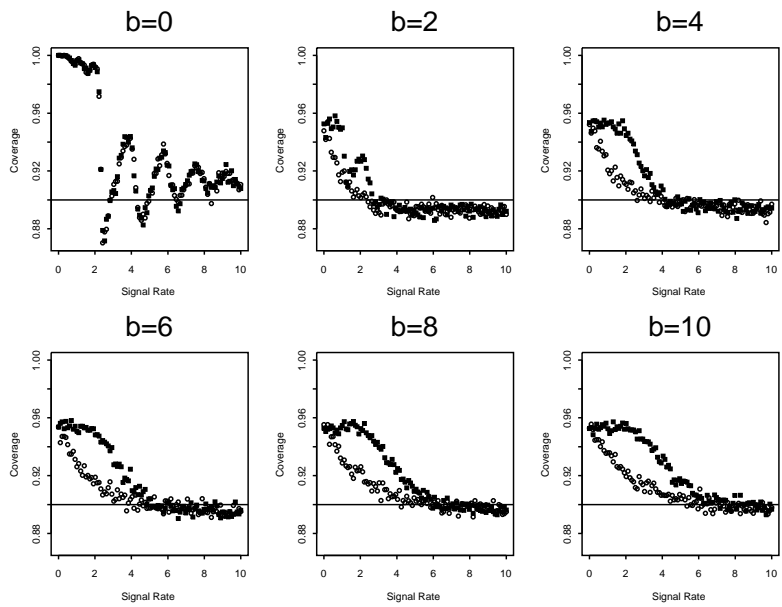}
	\caption{$90\%$ coverage
graphs when the signal is modeled as a Poisson and the background and the
efficiency are modeled as Gaussians with $\protect\sigma _{b}=0.5$, $e=0.85$
and $\protect\sigma _{e}=0.075$. The empty circles show the coverage using
the unbounded likelihood method and the solid squares show the coverage
using the bounded likelihood method.}
	\label{fig:fig5}
\end{figure}

\begin{figure}
	\centering
		\includegraphics[width=0.90\textwidth]{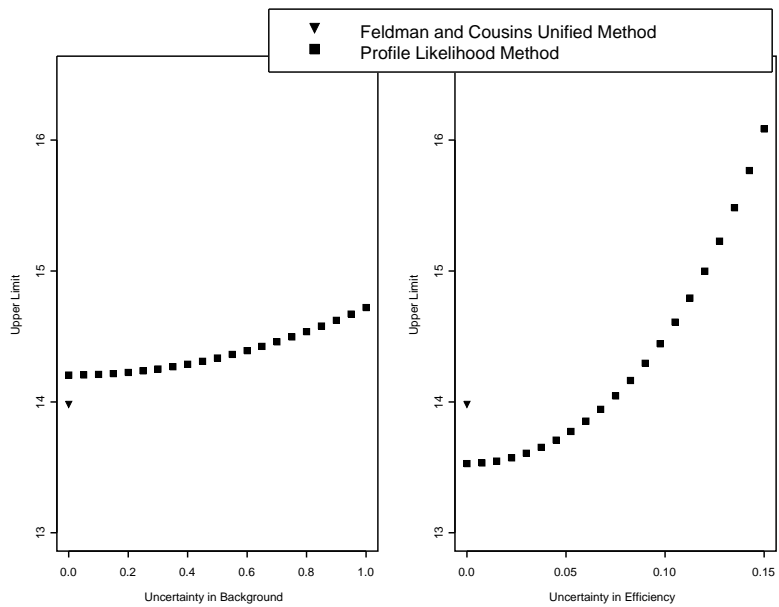}
	\caption{Upper limits as a
function of the uncertainties. In both graphs the background and the
efficiency are modeled as Gaussians. In the left panel we have the case $x=5$%
, $y=3$, $z=0.5$, $\protect\sigma _{e}=0.1$ and the uncertainty in the
background goes from $0.0$ to $1.0$. In the right panel we have the case $x=5
$, $y=3$, $z=0.5$, $\protect\sigma _{b}=0.75$ and the uncertainty in the
efficiency goes from $0.0$ to $0.15$. We have added the limits derived from
Feldman and Cousins unified method which ignores the uncertainties.}
	\label{fig:fig6}
\end{figure}

\end{document}